# *Polarization effect of zinc on the region 1-16 of amyloid-beta peptide: a molecular dynamics study*

Yan-Dong Huang[1] and Jian-Wei Shuai[1,2]

[1] Department of Physics and Institute of Theoretical Physics and Astrophysics, Xiamen University, Xiamen 361005, China
[2] Fujian Provincial Key Laboratory of Theoretical and Computational Chemistry, Xiamen University, Xiamen 361005, China

**Corresponding Author:**

Dr. Jianwei Shuai
Department of Physics
Xiamen University
Xiamen, Fujian 361005
China
Tel.: (86) 592-218-2575
Fax: (86) 592-218-9426
Email: jianweishuai@xmu.edu.cn

# Abstract

Zinc is found saturated in the deposited Amyloid-beta ($A\beta$) peptide plaques in Alzheimer's disease (AD) patients' brains. Zinc binding to $A\beta$ promotes aggregations, including the toxic soluble $A\beta$ species. Up to now, only the region 1-16 of $A\beta$ complexed with Zinc ($A\beta_{1-16}$-Zn ) is defined structurally in experiment, requiring an efficient theoretical method to present the interaction between zinc and $A\beta$ peptide. In order to explore the induced polarization effect on the global conformation fluctuations and the experimentally observed coordination mode of $A\beta_{1-16}$-Zn, in this work we consider an all-atom molecular dynamics (MD) of $A\beta_{1-16}$-Zn solvated in implicit water. In our model the polarization effect affects the whole peptide is applied. The induced dipoles are divided into three distinct scales according to their distances from zinc. Besides, the atomistic polarizability on the coordinating sidechains is rescaled to describe the electron redistribution effect. As a comparison, another model which exactly follows Sakharov and Lim's method (J. Am. Chem. Soc., 127, 13, 2005) has been discussed also. We show that, associated with proper van der Waals (vdW) parameters, our model not only obtains the reasonable coordinating configuration of zinc binding site, but also retains the global stabilization, especially the N-terminal region, of the $A\beta_{1-16}$-Zn. We suggest that it's the induced polarization effect that promotes reasonable solvent exposures of hydrophobic/hydrophilic residues regarding zinc-induced $A\beta$ aggregation.

## 1. Introduction

Alzheimer's disease (AD) is a fatal neurodegenerative disorder and characterized by aggregation of amyloid-beta peptide ( $A\beta$ ) into insoluble plaque in AD patient brains. Zinc is found with high concentration in the plaque [1]. It's believed that zinc involves in the event of $A\beta$ aggregation as a key factor for the pathogenesis of AD [2,3,4,5,6,7,8,9,10]. The original hypothesis for AD is the so-called amyloid cascade hypothesis [11]. But, in recent years, soluble $A\beta$ aggregates that are formed at the early stage of aggregating have been considered the pathogenic molecular form of $A\beta$ [8,12]. Further more, it is revealed that not only the size of $A\beta$ oligomer but also the peptide conformation define the toxicity of $A\beta$-Zn complexes [8,9,13,14,].

As a consequence, it's of interest to study the role of zinc in such pathological oligomerizations of $A\beta$. Nevertheless, biophysical studies of the structure of synthetic $A\beta$ are subjected to many difficulties. One important problem is related to aging-related aggregation of $A\beta$ in water solution [15,16]. As a result, it's difficult to experimentally obtain the structural information of pathogentic $A\beta$ oligomers forming spontaneously or promoted by metal ions. At present, there is still not any experimental evidence to show the structural information of $A\beta$-Zn complexes, calling for the prediction with theoretical simulation.

Considering an accurate description of zinc binding site interactions, the first-principal MD method or hybrid quantum mechanics/molecular mechanics MD method (QM/MM MD) has been recently used to study the local/global conformation fluctuations of truncated $A\beta$-Zn complex [16,17]. Nevertheless, $A\beta$ has been demonstrated to be an intrinsically disorder peptide [18,19,20], so it's not sufficient to account for the impact of zinc on $A\beta$ in a local domain or small timescale. For example, the 11$^{th}$ amino acid (a. a.) residue (Glu11) of $A\beta$ actually displays a bidentate binding fashion [21]. But the QM-based models suggest a monodentate binding to zinc [16,17], which might due to the lack of correct description of the protein environment. In order to improve the quality of conformational sampling, more extensive interactions between zinc and $A\beta$ should be taken into account. However, computational studies of the biomolecular structures in the framework of quantum mechanics schemes are often hampered by extremely demanding computer time, though, in principle, they provide the most accurate structures.

On the other hand, classical molecular dynamics methods with a variety of force fields have been developed and widely used to study the global/local conformational fluctuations of zinc proteins in a larger timescale with more extensive samplings [22,23,24,25,26,27]. In physical view of point, it's not accurate by neglecting the redistribution of electrons on the amino acid residues that bind to zinc directly. This is because the electrons in several residue sidechains, such as imidazole ring in histidine and benzene rings in tryptophan or phenylalanine, are highly delocalized under strong electric field created by zinc. As a result, binding of zinc can not only induce atomic point dipoles, but also result in dramatic polarization of polarizable molecular fragments [28,29,30]. Besides, due to the binding of zinc, the out-shell electrons of negative ligating ions, like deprotonated nitrogen in histidine and deprotonated sulfur in cysteine, have a certain probability on the orbital of zinc. As a consequence, charges transfer from ligating atoms to zinc. So, the two quantum effects, i.e. the polarization effect and the charge transfer effect, are critical for MD simulation of $A\beta$-Zn complex [29,30,31].

However, most of the current empirical force fields do not explicitly incorporate effects of zinc induced polarization and charge transfer. These models use the means of force field to describe the induced interactions of the zinc with water molecules and proteins atoms. In recent years, the impact of zinc on the conformation of $A\beta$ monomer or oligomer has been studied based on the MD simulations [32,33,34]. However, in these studies, zinc-ligand distances are constrained to their NMR evaluated distances, which would hinder global sampling of conformations and further exploration of kinetic process involved in zinc-coupled aggregating of $A\beta$.

On the basis of the nonbonded model presented by Stote and Karplus [22], Sakharov and Lim additionally consider two induction effects: local polarization effect presented with atom-based point dipole model and charge transfer effect specified by updating charges of coordinating atoms as well as zinc at any time step during simulations [29,35]. As a complete description of atom-atom interaction, their potential energy function includes all energy components derived from the well known Kitaura-Morokuma energy decomposition [36]. For example, the 6-12 L-J potential contains exchange and dispersion components; the coulombic potential with variable charges includes the induced charge transfer effect as well as the point charge electrostatic energy; and the polarization potential represents the induced polarization effect. By incorporating the two quantum effects (i.e., induced polarization and charge transfer) empirically into the conservational nonbonded potential energy function and combining with appropriate vdW parameters, they successfully captured experimentally observed coordinating structures of solely structural zinc binding sites of zinc-finger proteins [29].

Sakharov and Lim presented the induced dipoles at atom-based level [29]. Their model, in practice, might hinder its implementation into $A\beta$-Zn like complexes where polarization effect is no longer confined to the first-shell ligands. The calculation of polarization field on the whole peptide is bound to increase computational cost significantly.

It's worth to mention that most zinc binding sites can be classified to be structural and catalytic [37,38]. According to a statistical survey of zinc proteins, if there is no more than one Cys out of the ligating a. a. residues, the zinc binding site is defined to be catalytic [37]. Thus, accordingly the zinc binding site in $A\beta$, including three Histidines and one Glutamic Acid, belongs to the catalytic site. In most catalytic sites, ligands transfer less amount of charges than that in structural sites, enabling zinc to serve as a lewis acid. As a consequence, the charge-dependent polarization capacity of zinc in catalytic sites is generally stronger than in structural sites [37]. Therefore, it's quite necessary to account for the global impact of zinc-induced polarization force on $A\beta$. In fact, it's noted that those residues (i.e. Arg5 and Gly9 in $A\beta$) relatively far from zinc are also polarized [16].

The NMR structures of isolated $A\beta$ monomer or fragments are accessible [21,39,40,41,42,43], and the structure of 1-16 region of $A\beta$ complexed with zinc ($A\beta_{1-16}$-Zn) has also been defined in NMR experiment by Zirah et al. [21]. Recently it has been shown that the minimal zinc binding domain of $A\beta$ is the 6-14 region of $A\beta$ ($A\beta_{6-14}$-Zn) [17]. Therefore, in the present work, we adopt the coordinate sets of $A\beta_{1-16}$-Zn from NMR experiment to discuss the $A\beta_{1-16}$-Zn complex with an atomistic MD simulation.

Different from the previous strategies applied within CHARMM22 force field and outlined by Sakharov and Lim [29], we consider the zinc-induced dipoles at three resolutions, ranging from the residue-based level, to the intermediate level, and to the atom-based level. With atom-based dipoles to the zinc binding site, some atomic information, such as the coordination bond length and number, are kept and used in the description of the zinc binding site. While with the multiscale dipoles to the whole paptide, the amount of computation increases linearly only with the amino acid residue number. Furthermore, the induced polarization force acts on the whole peptide, making it more reasonable to study the solvent exposure of hydrophobic/hydrophilic residues.

For the charge transfer effect, a linear approximation is employed here to calculate the distance-dependent charge transfer from each ligating atom to zinc. Different from the symmetrical coordination configuration in zinc-finger protein discussed in Sakharov and Lim's model [29], the coordinating configuration of $A\beta_{1-16}$-Zn is asymmetrical**.** As a result, in the preliminary QM calculations, the amount of charge transferred from each ligating atom to zinc has to be calculated individually.

In this work, our MD method has been assessed from some statistic properties, such as root-mean-square deviations from the experimental structure, the radius of gyration of $A\beta_{1-16}$-Zn, the coordinating bond length between metal and ligating atoms, and hydrogen bonds involved in the secondary structure of interest. In addition, solvent-accessible-surface area of individual residue is also recorded to explore solvent exposure of a. a. residues. In the end, our method is applied to another zinc complex with also catalytic-type zinc binding site. The results show that our method is able to reproduce the coordinating configuration as well as the global stabilization of $A\beta_{1-16}$-Zn, which is compatible with experimentally determined structures. Besides, our method can be applied to other $A\beta_{1-16}$-Zn like complexes.

## 2. Methods

In this work, the potential energy function (Eq. 1) that describes the interaction between zinc and $A\beta_{1-16}$ peptide consists of electrostatic polarization force, the Coulum force and vdW force, which is given by

$$V_{Zn,\mathrm{Protein}} = V^{pol} + \sum_i \left\{ \frac{(q_{Zn}^0 + \sum_i \Delta q_{i\to Zn})(q_i^0 - \Delta q_{i\to Zn})}{4\pi\varepsilon_0 r_{Zn,i}} + 4\varepsilon_{Zn,i}\left[\left(\frac{R_{Zn,i}}{r_{Zn,i}}\right)^{12} - 2\left(\frac{R_{Zn,i}}{r_{Zn,i}}\right)^{6}\right]\right\}$$

(1)

where the formula is similar with the conservational nonbonded model [22], expect for an additional term, electrostatic polarization energy $V^{pol}$. The initial charges, $q_{Zn}^0$ and $q_i^0$, of zinc and peptide atoms are extracted from CHARMM22 force field. The charge transfer effect is incorporated by adjusting charges on zinc and its coordinating atoms at any time step during simulations. $\Delta q_{i\to Zn}$ is the charge transferred from the *i*-th atom to zinc, while $\sum_i \Delta q_{i\to Zn}$ the summation of charges transferred from protein atoms to zinc. In practice, only those atoms that directly form coordinating bonds with zinc donate charges.

## 2.1 Charge transfer estimations

We estimate the amount of charges transferred from the four amino acids (His6, Glu11, His13 and His14) of $A\beta_{1-16}$ to zinc. Three neutral histidines and one negatively charged glutamic acid are modeled respectively by three imidazole rings (IMI) and one formate ( $HCOO^-$ ) that tetra-coordinated to zinc ($[Zn^{2+}(IMI)_3HCOO^-]^{1+}$). The initial structure of complex $[Zn^{2+}(IMI)_3HCOO^-]^{1+}$ is extracted from the NMR structure (PDB code: 1ZE9) [21] and fully optimized at the $B3LYP/6-31+G^*$ level using the Gaussian 03 program [44]. For the optimized complex, the transferred charges are then calculated by using the NBO Version 3.1 module in Gaussian 03 at the same level, where natural population analysis and the second order perturbation theory analysis of fock matrix in NBO basis are considered.

Assuming a highly symmetric structure between Zinc and ligands, Sakharov and Lim suggested that each ligand in the complex transfers the same amount of charge to the zinc. Thus they calculated the amount of charge donated by each ligand simply from the charge on zinc [29]. Differently, the geometry of $[Zn^{2+}(IMI)_3 HCOO^-]^{1+}$ complex has lower symmetry. So we have to explicitly calculate the charge donated from each ligating atom to zinc. Based on the values of second order perturbation energy in NBO calculation, the amount of charge accepted by zinc appears to be contributed mainly by the coordinating atoms. While, the $\pi$ orbital metal-to-ligand donation can be excluded because the ligand atoms are all poor electron acceptors [45]. Therefore here we focus on the charges donated by the atoms that directly bind to zinc. The transferred charge is then given by

$$\Delta q_{L \to Zn} = \left[F(L, Zn) / (E_{Zn} - E_L)\right]^2 \tag{2}$$

where $\Delta q_{L \to Zn}$ is the quantity of charge transferred from the lone pair orbital (LP) of the donated atoms to the anti-LP orbital (LP$^*$) of zinc, $F(L, Zn)$ presents the Fock matrix, and $E_{Zn} - E_L$ is the energy difference between LP* of zinc and LP of ligand atom *L*.

The three ligating nitrogens in imidazoles and the ligating oxygen in carboxylate respectively donate 0.094e, 0.094e, 0.084e and 0.13e, the sum of which occupies 98.5% of the amount of charge, 0.41e, that zinc accepts. The equilibrium distances between zinc and the three ligating nitrogens and one oxygen are 2.05 Å, 2.05 Å, 2.08 Å and 1.98 Å, respectively. As implemented in the model by Sakharov and Lim [29], the amount of charge $\Delta q_{L \to Zn}$ transferred from the ligand atom *L* to zinc is linearly related with their interatomic distance $r_{Zn,L}$, as given by

$$\Delta q_{L \to Zn} = a_L \times r_{Zn,L} + b_L \tag{3}$$

in which the atom *L* represents the ligating nitrogens in the imidazole and the ligating oxygen in the formate. The parameters $b_N$ and $a_N$ for nitrogens can be obtained based on the average equilibrium distance $\bar{r}_{Zn,N} = (2.08+2.05+2.05)/3 = 2.06$ Å, the transferred charge $\overline{\Delta q}_{N\to Zn} = (0.094+0.094+0.084)/3 = 0.091e$, and the vdW radius. $b_O$ and $a_O$ for the ligating oxygen can be calculated in a similar fashion. At any time-step $t$ during simulations, the transferred charge $\Delta q_{L\to Zn}$ is changed according to Eq. 3 only if $r_{Zn,L}$ is smaller than the sum of the vdW radii of zinc $R_{Zn}$ and the ligating atoms.

In the paper, two sets of vdW parameters are discussed. As shown in Tab. 1, the vdW2 is the vdW parameter set of zinc from CHARMM22 force field [46], which can reproduce the experimental first-shell Zn-O distances, coordinating numbers (CNs) in water, and the absolute experimental hydration free energies of zinc. The vdW1 is the vdW parameter set of zinc obtained by adjusting vdW2 to reproduce the experimental hydration free energy of zinc relative to the experimental hydration free energy of other metal dications [47]. Thus, two sets of $a_L$ and $b_L$ are determined, as given in Tab. 1.

Table 1

| | $R_{Zn}$ (Å) | $\varepsilon_{Zn}$ (kcal/mol) | $a_N$ (e/Å) | $b_N$ (e) | $a_O$ (e/Å) | $b_O$ (e) |
|---|---|---|---|---|---|---|
| vdW1 | 0.88 | -0.183 | -0.14 | 0.38 | -0.22 | 0.57 |
| vdW2 | 1.09 | -0.25 | -0.1 | 0.3 | -0.16 | 0.45 |

## 2.2 Multiscale point dipole model

In our model, we describe the dipoles with three scales, ranging from the residue-based level, to the backbone/sidechain-based levle, and to the atom-based level. In the zinc binding site, If it is the backbone (or sidechain) of a residue that binds to zinc, then the atom-based dipole is considered for the backbone (or sidechain), and the polarization effect on the corresponding sidechain (or backbone) is simulated with a single coarse-grained dipole. The residue-based dipoles are considered for the residues that locate outside the zinc binding site. In detail, from the crystal structures obtained with NMR experiment, zinc tetrahedrally binds to the sidechains of residues Glu11, His6, His13 and His14 [21]. Thus, the dipoles are atom-based for the four coordinating sidechains (see the red groups in Fig. 1). Correspondingly, four backbone-based dipoles are applied to the backbones of these four residues (see the green ball in Fig. 1). For the rest 12 amino acid residues, residue-based dipoles are defined (see the white ball in Fig. 1).

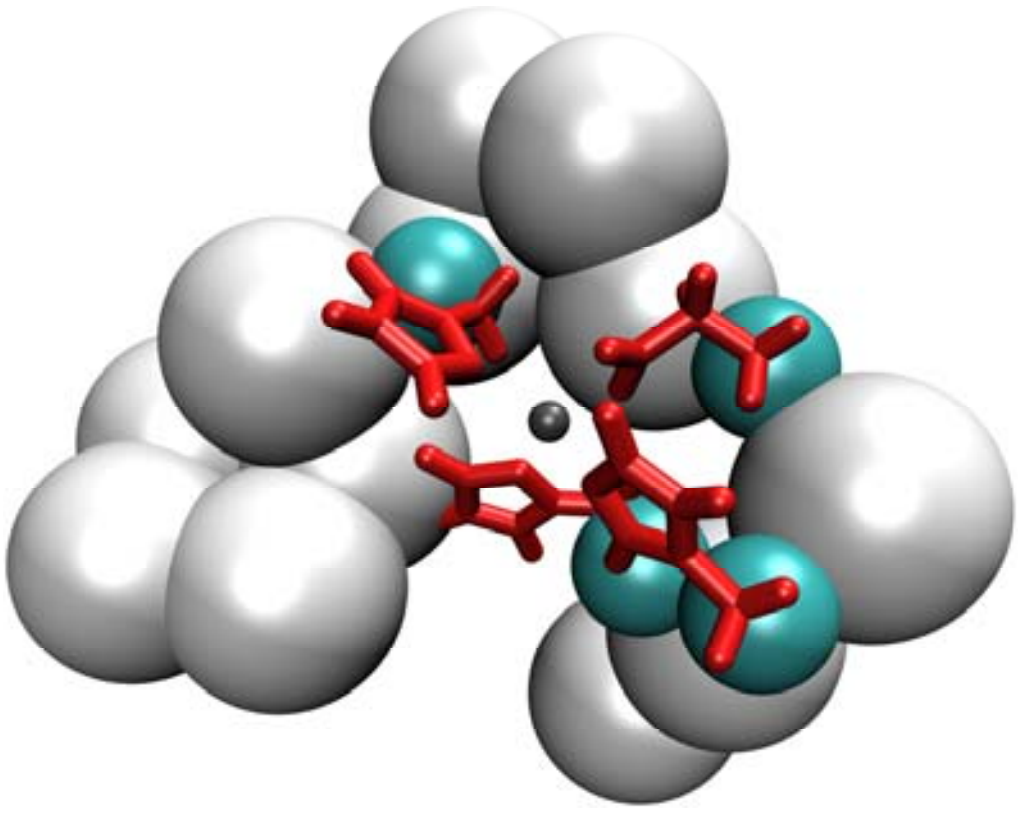

Figure 1. Schematic diagram of multiscale dipoles applied to the $A\beta_{1-16}$-Zn complex. The white ball denotes the residue-based dipole. The smaller green ball represents the backbone-based dipole. The dipoles in red color are the atom-based dipoles. The gray ball in the center is zinc.

The peptide atom-atom polarization interaction is much weak compared with the polarization effect induced by zinc, and has been implicitly included in CHARMM22 force field simply by adjusting atomic charges to reproduce the interaction energy [46]. Thus, in the present work, we follow the assumption in [29] that the biomolecule is polarized utterly due to the electric field created by the charge on zinc. As a result, the polarization energy $V^{P}$ is contributed by the electric interaction between all zinc-induced dipoles of the peptide and the point charge at zinc and the interaction between induced dipole at zinc and the point charges of peptide atoms:

$$V^P = V^A + (V_{q,\mu}^{PA} + V_{\mu,\mu}^{PA}) = -\frac{1}{2}\sum_i \vec{E}_i^0 \cdot \vec{\mu}_i \tag{4}$$

where

$$V^A = \frac{1}{2}\sum_i (\vec{E}_i^0 + \vec{E}_i^\mu) \cdot \vec{\mu}_i, \tag{5}$$

$$V_{q,\mu}^{PA} = -\sum_i \vec{E}_i^0 \cdot \vec{\mu}_i \tag{6}$$

and

$$V_{\mu,\mu}^{PA} = -\frac{1}{2}\sum_i \vec{E}_i^\mu \cdot \vec{\mu}_i, \tag{7}$$

in which, the activation energy, $V^A$, required to create the induced dipoles is known as the self-polarization energy [48]; $V_{q,\mu}^{PA}$ denotes the total interaction energy between all interacting charges and induced dipoles; $V_{\mu,\mu}^{PA}$ denotes the total interaction energy among all interacting induced dipoles. The sum of $V_{q,\mu}^{PA}$ and $V_{\mu,\mu}^{PA}$ is the polarization energy relative to $V^A$. Consequently, the net change in interaction energy, $V^P$, is obtained in equation 4. The summations in Eqs. 4-7 are over all induced dipoles. $\vec{\mu}_i$ is the dipole moment induced at the *i*-th induced dipole. $\vec{E}_i^0$ and $\vec{E}_i^\mu$ are the local electric fields at the *i*-th induced dipole created by neighboring point charges and induced dipoles, respectively.

Assuming a linear response approximation, the induced dipole moment $\vec{\mu}_i$ at the *i*-th site is proportional to the local electric field, $\vec{E}_i$ which is the sum of $\vec{E}_i^0$ and $\vec{E}_i^\mu$:

$$\vec{\mu}_i = \alpha_i \vec{E}_i \tag{8}$$

where the proportionality constant $\alpha_i$ is the polarizability of the *i*-th polariable unit. The standard polarizabilities of peptide residues and atoms are taken from Ref. [49]

and Ref. [50], respectively. The polarizability of zinc, 2.294 Å$^3$, is from Ref. [51]. And the polarizability of the deprotonated nitrogen on an imidazole, 2.8 Å$^3$, is from Ref. [29], which is bigger than the standard value 1.09 Å$^3$ provided in Ref. [50].

In fact, there are two factors that can affect the polarizabilities of polarizable atoms, namely the amount of electrons and the distance of the electrons from nuclear charge. The polarizability of an atom increases as the amount or the distance increases, because the nuclear charge has less control on charge distribution. Electrons on a ligating sidechain are delocalized and redistributed under strong electric field created by zinc. The redistribution of electrons leads to the increased polarizabilities of atoms that directly bind to zinc and the decreased polarizability of other atoms on the same sidechain. Here, the polarizabilities of the ligating atoms are adjusted by multiplying the standard values with a parameter 2.8/1.09, while the atomic polarizabilities of other atoms on the same ligating sidechain are modified simply by setting them zero. As to the fragments out of the zinc binding site, we assume that the electric field created by zinc is able to distort the electron cloud of atoms, but is not strong enough to affect the molecular orbital. Thus the residual polarizability has not to be modified.

The total electric field $\vec{E}_{Zn}$ at zinc is the vector sum of the electric field due to the current charges of protein atoms and induced dipoles in protein:

$$\vec{E}_{Zn} = E^0_{Zn} + \sum_i \vec{T}_{Zn,i}\vec{\mu}_i = \sum_j \frac{q_j \vec{r}_{Zn,j}}{r^3_{Zn,j}} + \sum_i \frac{\vec{\mu}_i}{r^3_{Zn,i}}\left(\frac{3\vec{r}_{Zn,i}\vec{r}_{Zn,i}}{r^2_{Zn,i}} - 1\right) \quad (9)$$

where $\vec{E}^0_{Zn}$ is the electric field created from the current charges of protein atoms with summation over all peptide atoms, and $\vec{T}_{Zn,i}$ is the *i*-th element of the dipole field tensor with summation over all the induced dipoles in the peptide.

The total electric field $\vec{E}_i$ at the *i*-th dipole in the peptide is the sum of the field $\vec{E}^0_i$ which is caused by the current charge and the field $\vec{T}_{i,Zn}\vec{\mu}_i$ which is caused by the induced dipole on zinc:

$$\vec{E}_i = E_i^0 + \vec{T}_{i,Zn}\vec{\mu}_i = \frac{q_{Zn}\vec{r}_{i,Zn}}{r_{i,Zn}^3} + \frac{\vec{\mu}_{Zn}}{r_{i,Zn}^3}\left(\frac{3\vec{r}_{i,Zn}\vec{r}_{i,Zn}}{r_{i,Zn}^2} - 1\right) \quad (10)$$

In turn, the corresponding induced dipole moments are obtained with Eq. 8. Subsequently, we can solve the self-consistent field by iterating the coupled equations (i.e. Eqs. 8-10) until the convergence condition that the polarization energy change is smaller than a minimal is satisfied.

In order to avoid the calculation of derivative of the polarization energy with respect to the induced dipole moments, the polarization energy is represented by another form here. Multiplying vector $\vec{\mu}_i$ on both side of Eq. 10, we can obtain the following identity:

$$-\frac{1}{2}\sum_i \vec{\mu}_i \cdot \vec{E}_i^q - \frac{1}{2}\sum_i \vec{\mu}_i T_{i,Zn}\vec{\mu}_{Zn} + \frac{1}{2}\sum_i \vec{\mu}_i \cdot \alpha_i^{-1}\vec{\mu}_i = 0 \quad (11)$$

As a result, we rewrite the induced polarization energy in the following form [52]:

$$V^{pol} = -\sum_i \vec{\mu}_i \cdot \vec{E}_i^q - \frac{1}{2}\sum_i \sum_{j \neq i} \vec{\mu}_i T_{i,j}\vec{\mu}_j + \frac{1}{2}\sum_i \vec{\mu}_i \cdot \alpha_i^{-1}\vec{\mu}_i \quad (12)$$

Since the derivative with respect to any induced dipole moment is zero, the polarization forces can be derived as follows:

$$\vec{f}_{Zn} = (\vec{\mu}_{Zn} \cdot \nabla_{Zn})\vec{E}_{Zn}^q + \sum_i [(\vec{\mu}_i \cdot \nabla_{Zn})\vec{E}_i^q + \frac{1}{2}(\vec{\mu}_i \cdot \nabla_{Zn})T_{i,Zn}\vec{\mu}_{Zn} + \frac{1}{2}(\vec{\mu}_{Zn} \cdot \nabla_{Zn})T_{Zn,i}\vec{\mu}_i] \quad (13)$$

$$\vec{f}_i = (\vec{\mu}_i \cdot \nabla_i)\vec{E}_i^q + (\vec{\mu}_{Zn} \cdot \nabla_i)\vec{E}_{Zn}^q + \frac{1}{2}(\vec{\mu}_{Zn} \cdot \nabla_i)T_{Zn,i}\vec{\mu}_i + \frac{1}{2}(\vec{\mu}_i \cdot \nabla_i)T_{i,Zn}\vec{\mu}_{Zn} \quad (14)$$

The approximation used in Eq. 8 shows a growth of the induced dipoles with the distances between induced dipoles and zinc, which becomes no longer valid at close distances. In order to avoid unphysical growth of the induced dipoles at close distance, a cutoff distance $r_{i,Zn}^{cutoff}$ is considered here, which is assumed to be equal to the sum of the vdW radii of atoms (residues) and zinc, and then multiplied by the parameter (=0.92), as suggested in [29]. For each dipole-dipole pair, $r_{i,Zn} = r_{i,Zn}^{cutoff}$ if $r_{i,Zn} \le r_{i,Zn}^{cutoff}$.

The center of a coarse-grained dipole is the centroid of the corresponding molecular fragment (residue, backbone or sidechain in the peptide). Thus, the polarization force vector $\vec{F}^{pol}$ that goes through the dipole center does not cause rotating motion of the fragment. As a result, the polarization force $\vec{f}_i^{pol}$ acting on atoms that belong to the fragment can be derived simply by the following equation

$$\vec{f}_i^{pol} = m_i \frac{\vec{F}^{pol}}{\sum_i m_i} \tag{15}$$

where $m_i$ is the atomic mass and the summation is over the atoms in the fragment.

## 2.3 Simulation Methodology

The starting structure of $A\beta_{1-16}$-Zn is the best representative conformer among 20 NMR structures. As to the atom-atom interaction between peptide atoms, we employ the popular all-atom CHARMM22 force field in conjunction with the CMAP extension [46, 53 ]. CHARMM22 is parameterized in reproducing protein conformational distributions in MD simulations. CMAP is a 2D dihedral energy correction map to the CHARMM22 and used to improve the sampling of backbone torsion angles. Here, the CHARMM22/CMAP is called the traditional force field. The MD simulations in either explicit or implicit solvent have been carried out in NVT ensembles under the experimental temperature of 278K. All simulations are carried out near neutral pH, so the acid residues such as Glu and Asp are deprotonated, whereas base resiudes like Arg and Lys are protonated.

Explicit solvent MD simulations with standard software are performed under the traditional force field. On the other hand, implicit solvent MD simulations are performed under the new force field where charge transfer and induced polarization effect are added to the traditional force field .

### 2.3.1 Explicit Solvent Simulations

The MD simulation software NAMD is used to simulate the $A\beta_{1-16}$-Zn complex solvated in a water box filled with TIP3P water molecules [54]. Counterions are not needed for the neutral $A\beta_{1-16}$-Zn complex. The vdW and electrostatic forces are switched at a distance of 10.0 Å to zero at 12.0 Å. Langevin thermostat is used to keep temperature constant and the viscosity is $10\,ps^{-1}$ [55,56]. The time step is 1fs. In order to eliminate close contacts, firstly the solute $A\beta_{1-16}$-Zn is fixed, while the solvent is relaxed at an NVE ensemble until the water reaches equilibrium. Secondly, the solute is released, and the whole system undergoes 50ps of energy minimization. Finally, a 30ns simulation is carried out.

### 2.3.2 Implicit Solvent Simulations

Simulations with explicit solvent molecules require prohibitive cost of computer resource and time especially in solvent relaxing. Implicit solvent has been proved to be more efficient than explicit solvent for small peptide complex like $A\beta_{1-16}$-Zn. Thus, in order to speed up the simulation and meanwhile to improve the efficiency of sampling with a short simulation, we employ the Generalized Born/Solvent Accessible Surface Area (GB/SASA) model [57,58,59] as an implicit solvent model. GB method approximates the polar contribution to the solvation free energy, while nonpolar contribution can be approximated with a term proportional to the SASA [60].

The GB model adopted here is originally proposed by Still and co-workers [61] and parameterized later specifically for proteins, peptides and nucleic acids within the CHARMM all hydrogen and polar hydrogen force fields [57]. Comparing with the standard numerical values derived from finite differential Poission-Boltzmann (FDPB) method implemented in software Delphi [62], the vdW radius of zinc as the intrinsic radius in GB model would produce underestimated effective Born radius of zinc and therefore overestimated solvation free energy. In order to overcome these problems, following the approach applied to the intrinsic radius of polar hydrogen [57,61,63], we set the intrinsic radius of zinc to 1.8 Å.

SASA values are obtained with the LCPO approximation [58] with a solvent probe radius of 1.4 Å. Atomic SASAs are computed by using

$$A_i = P_1 S_1 + P_2 \sum_{j\in N(i)} A_{ij} + P_3 \sum_{\substack{j,k\in N(i)\\ k\in N(j)\\ k\neq j}} A_{jk} + P_4 \sum_{j\in N(i)} A_{ij} \left( \sum_{\substack{j,k\in N[i]\\ k\in N(j)\\ k\neq j}} A_{jk} \right) \tag{16}$$

where $N(i)$ denotes the neighbor list (NL) of $i$ (the list of neighboring spheres that overlap with sphere $i$). In the first term, $S_1$ is the surface area of the isolated sphere corresponding to atom $i$. $A_{ij}$ is the area of sphere $i$ buried inside sphere $j$. Therefore, the second term involves the sum of pair-wise overlaps of sphere $i$ with its neighbors. The third term presents the sum of overlaps of neighbors of $i$ with each other. The fourth term is a further correction for multiple overlaps. Besides, in order to further improve computational efficiency, an optimization method called neighbor-list reduction (NLR) [59] is employed when SASA is computed, which allows selected neighbors of a central atom to be removed from the computation in a preprocessing step. Thus it allows the calculation of the atom's surface area to proceed with a shorter list of neighbors [59]. In the current work, new parameters *P*1 to *P*4 with respect to different types of atoms for both LCPO and LCPO/NLR are derived using our own training set of zinc proteins so as to make them consistent with CHARMM22 force field (See the supporting information). Giving the total SASA of the solute, nonpolar solvation free energy is equal to SASA timing a phenomenological surface tension coefficient of 0.00542 $kcal/(mol \cdot A^2)$.

The implicit solvent simulations are also combined with Langevin dynamics with the same viscosity 10 $ps^{-1}$. Notably, the external friction and random forces here act only on the heavy atoms on the peptide surface to avoid unphysical fluctuations of buried atoms. Time step is set at 1fs too, but cut-off approach is not used. Similarly, in order to avoid close contacts, the energy of the complex is minimized for 50ps before a 15ns simulation is carried out.

# 3. Results

Mainly two MD simulations of $A\beta_{1-16}$-Zn complex solvated in the implicit solvent are carried out in the current work. The first simulation (Sim1) adopts the model proposed by Sakharov and Lim [29] who consider only the atom-united dipoles for the sidechains that directly bind to zinc with the vdW parameter set of vdW1 (see Tab. 1). The second simulation (Sim2) corresponds to our model, in which the multiscale dipoles are used to account for the polarization effect of zinc on whole peptide. Sim2 adopts the vdW parameter set of vdW2 (see Tab. 1). In order to compare the results produced by Sim1 and Sim2, some structural and dynamic quantities are computed by averaging along the trajectories. Images are rendered with VMD [64]

## 3.1 Root-mean-square Deviation (RMSD)

In the simulation, the initial positions of atoms are determined from the NMR structure. Because in the NMR structure, the first and the last residues are weakly defined, the RMSD of the backbone Ca atoms is then calculated without these two residues. The RMSD trajectory for $A\beta_{1-16}$ is given by the blue line in Fig. 2. The RMSD trajectory of the backbone Ca atoms for the minimal zinc binding domain, $A\beta_{6-14}$, including all four coordinating a. a. residues, is also calculated (green line in Fig. 2). We also calculate the RMSD trajectory of zinc binding center, containing zinc and another five atoms that directly binds to it (red line in Fig. 2).

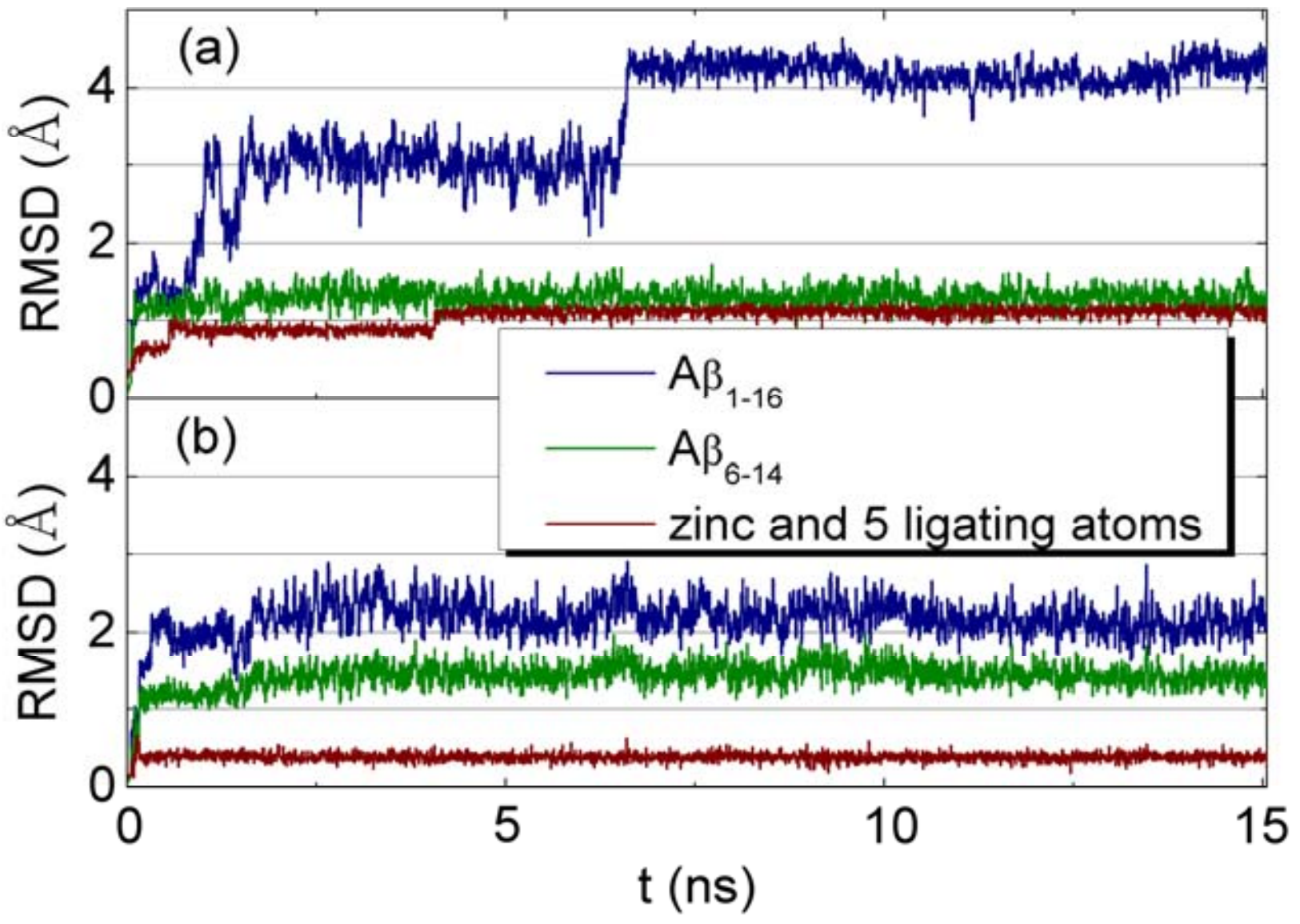

**Figure 2**. RMSD trajectories as a function of time derived from (a) Sim1 and (b) Sim2. The RMSDs of the backbone Ca atoms for the whole peptide $A\beta_{1-16}$ and for minimal zinc binding domain $A\beta_{6-14}$ are represented with blue and green lines respectively. The RMSD of the zinc binding center, including zinc and another five atoms that directly bind to zinc, is represented with red lines.

For Sim1 in Fig. 2a, the RMSD of $A\beta_{6-14}$ equilibrates fast and keeps stable throughout the simulation, but there are two evident jumps away from the native structure for the RMSDs of $A\beta_{1-16}$ and zinc binding center before they reach equilibrium. For Sim2 in Fig. 2b, one can see that all three RMSD trajectories keep stable after the short transient period.

Averaging the RMSD trajectories along the time after convergence (here and after we use the period of last 5ns as the sampling region), the mean as well as standard deviation of RMSD is obtained (see Tab. 2). One can see that either Sim1 or Sim2 is able to retain the structure of the minimal zinc binding domain (1.30 Å for Sim1 and 1.44 Å for Sim2), however, Sim2 produces much smaller RMSD values of whole $A\beta_{1-16}$ and RMSD values of zinc binding center. Smaller RMSD of $A\beta_{1-16}$ for Sim2 indicates the significance of induction effect of zinc on the structural stabilization of the global region of $A\beta_{1-16}$ peptide. Thus, compared to Sim1, Sim2 replicates better the conformational stabilization of the N-terminal region in the presence of zinc. The using of vdW2 as well as the rescaling of atomic polarizability to the atoms on ligating sidechains can stabilize better the structure of zinc binding center.

Table 2

| RMSD (Å) | Sim1 | Sim2 |
|---|---|---|
| $A\beta_{1-16}$ | $4.16 \pm 0.15$ | $2.14 \pm 0.17$ |
| $A\beta_{6-14}$ | $1.30 \pm 0.12$ | $1.44 \pm 0.12$ |
| zinc and 5 ligating atoms | $1.11 \pm 0.06$ | $0.39 \pm 0.05$ |

## 3.2 Radius of Gyration

Now we discuss the radius of gyration $R_g$ which is defined as

$$R_g = \sqrt{\frac{m_i(\vec{r}_i - \vec{r}_c)^2}{\sum_i^N m_i}} \qquad (17)$$

where $\vec{r}_c$ is the vector to specify the center of mass of the complex with *N* atoms, $\vec{r}_i$ is the position vector of atom *i* with mass $m_i$, and the sum is over the *N* atoms in the complex.. The radius of gyration as a function of time is illustrated in Fig. 3 where the blue dash line is the $R_g^0$ of the NMR structure. Compared with $R_g^0$, Sim2 provides a better $R_g$ than that produced by Sim1. It has been stressed in [21] that the structure of $A\beta_{1-16}$ would become more compact upon zinc binding. Thus, the bigger $R_g^0$ with higher fluctuations produced by Sim1 which is similar with that found in apo $A\beta_{1-16}$ implies that the impact of zinc is underestimated in Sim1. These results indicate that the global consideration of the zinc-induced dipoles is critical to keep the complex under a reasonable radius of gyration.

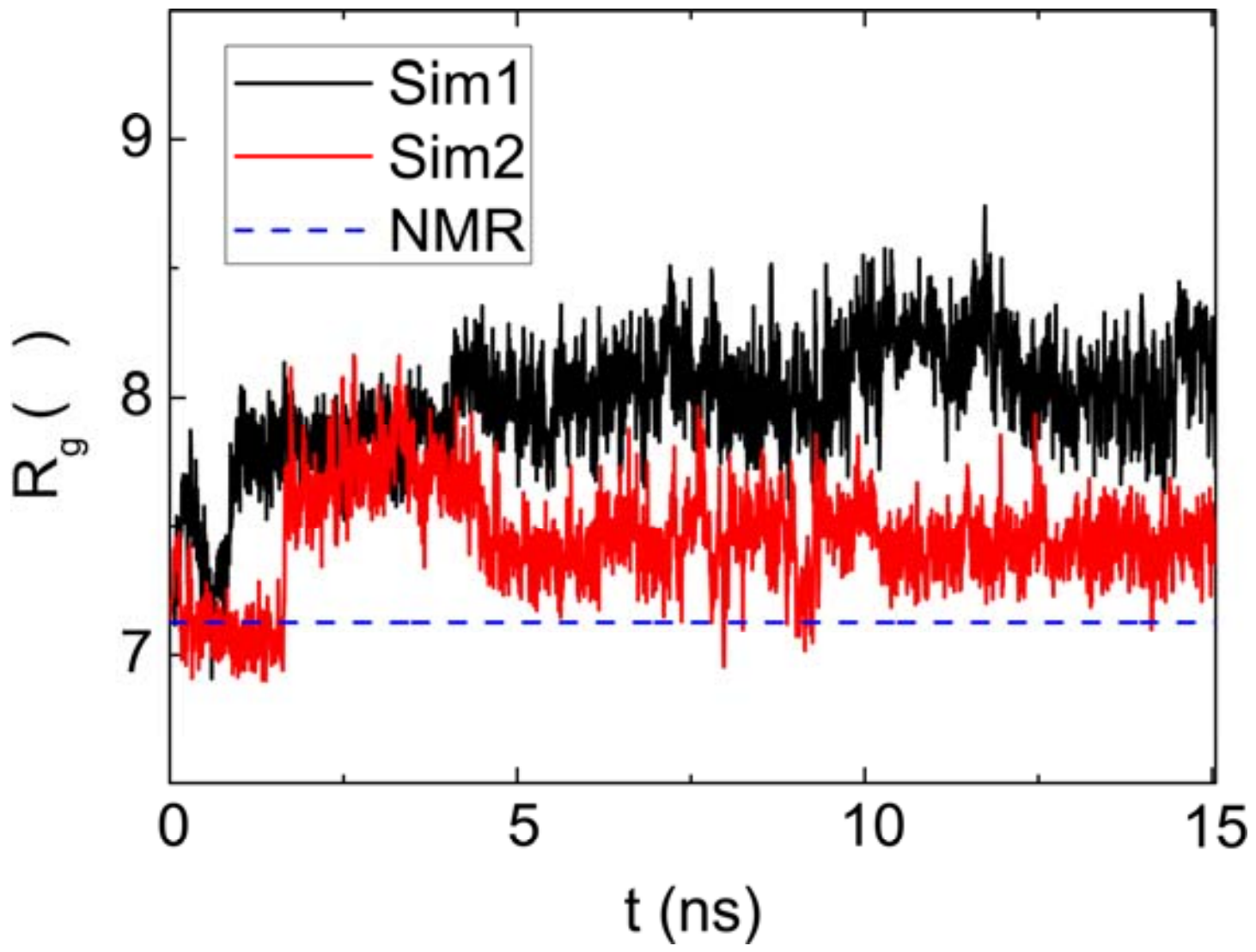

**Figure 3.** Trajectories of radius of gyration of complex $A\beta_{1-16}$-Zn as a function of time by Sim. 1 (black line) and by Sim2 (red line). The dash line is the standard value derived from the NMR structure.

## 3.3 Coordination geometry and secondary structure

In the NMR structure, zinc is tetrahedrally coordinated to four a. a. residues, namely Glu11 and His (6,13,14). The carboxylate sidechain of Glu11 binds to zinc in a bidentate fashion with the Zn-O distance 2.11 Å. $N\delta$ atoms of His6 and His14 have a distance of about 2.1 Å and 2.3 Å from zinc, respectively, while $N\varepsilon$ atom of His13 about 2.14 Å. As highlighted in Fig. 4b, both oxygens on the carboxylate coordinate with zinc. However, in Fig. 4a one of the two ligating oxygens is excluded off the zinc binding center. From Tab. 3, one can see that, compared to the NMR data, Sim2 gives better coordination bond lengths than Sim1. The mean absolute deviation of the bond lengths from the NMR data obtained by Sim2, i.e. 0.056 Å, is much smaller than 0.51 Å obtained by Sim1.

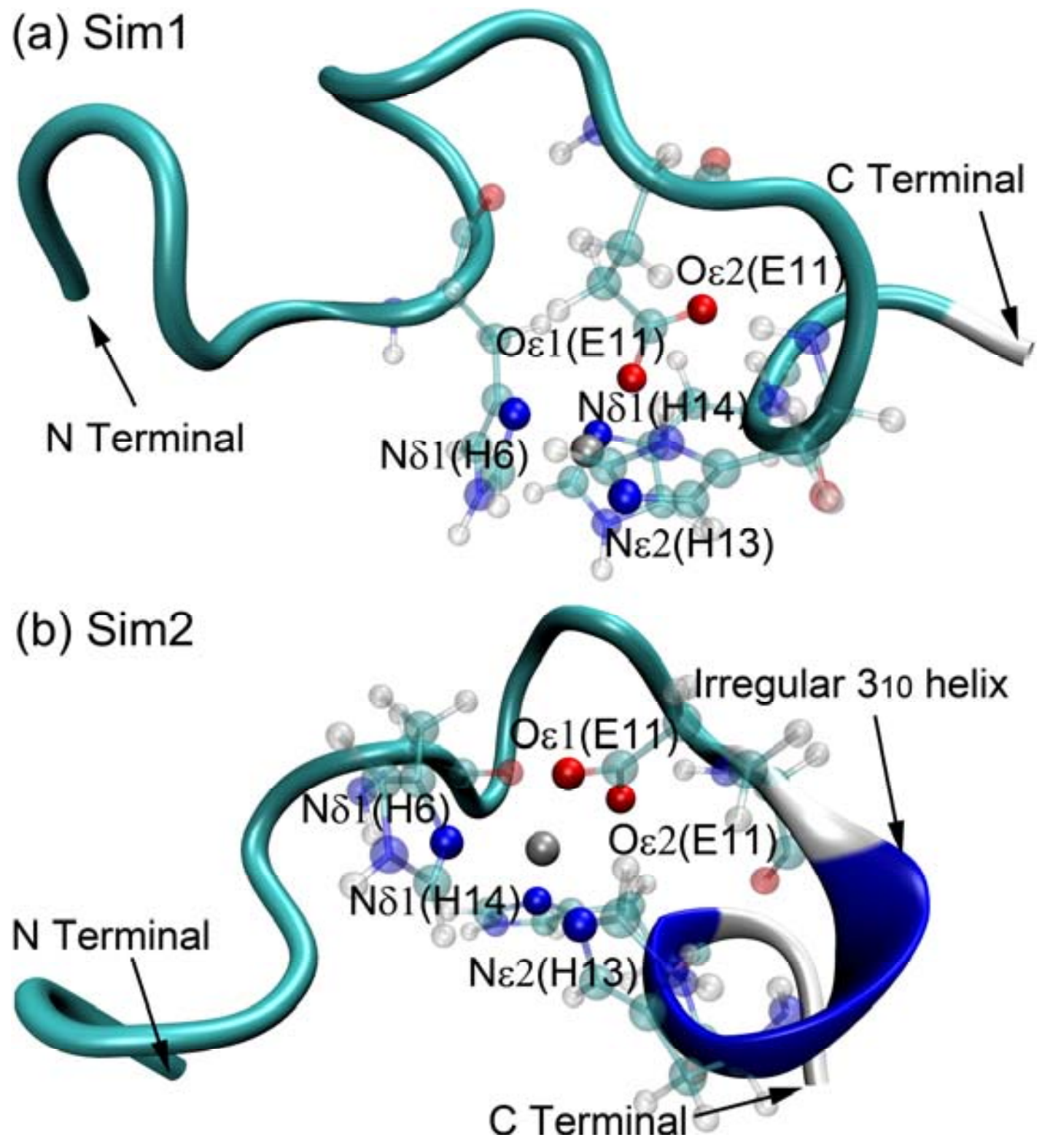

**Figure 4.** Snapshots of $A\beta_{1-16}$-Zn complex from MD simulations (a) Sim1 and (b) Sim2 respectively. The ligating residues (His(6, 13, 14) and Glu11) are illustrated with ball-bond mode. The coordinating atoms, including $N\delta 1$ (His6), $N\varepsilon 2$ (His13), $N\delta 1$ (His14), $O\varepsilon 1$ (Glu11) and $O\varepsilon 2$ (Glu11), are highlighted by making the rest atoms transparent. The abbreviations E and H denote Glu and His respectively. Carbon is in green color, hydrogen in white, oxygen in red and nitrogen in blue. The blue helix-like structure is the irregular $3_{10}$ helix.

The smaller coordination bond lengths derived from Sim1 are mainly due to the use of zinc radius 0.88 Å in vdW1, which was suggested in studying zinc protein [29]. There is an irregular $3_{10}$ helix in the C terminus of $A\beta_{1-16}$ characterized by two hydrogen bonds formed by backbone O (Glu11)- H-N (His14) (HBond1) and O (Val12)- H-N (Gln15) (HBond2) respectively. When the distance between donor (backbone O) and acceptor (backbone N) is smaller than 3.5 Å and simultaneously the hydrogen bond angle is bigger than 110$^o$, a hydrogen bond is considered to exist. The mean distances as well as angles of the two backbone hydrogen bonds are listed in Tab. 4. These results indicate that, although the HBond2 is captured, Sim1 fails in reproducing the HBond1 since the donor-acceptor distance 4.46 Å exceeds the up limit.

Table 3

| Coordination Bond Length (Å) | Sim1 | Sim2 | NMR |
|---|---|---|---|
| $N\delta$ (His6) | $1.96 \pm 0.04$ | $2.14 \pm 0.04$ | 2.11 |
| $N\varepsilon$ (His13) | $1.92 \pm 0.04$ | $2.16 \pm 0.04$ | 2.15 |
| $N\delta$ (His14) | $1.93 \pm 0.04$ | $2.14 \pm 0.04$ | 2.29 |
| $O\varepsilon_1$ (Glu11) | $1.92 \pm 0.06$ | $2.06 \pm 0.04$ | 2.11 |
| $O\varepsilon_2$ (Glu11) | $3.83 \pm 0.09$ | $2.07 \pm 0.04$ | 2.11 |

Table 4

| | Hydrogen bond angle (Deg.) | | Donor-acceptor distance (Å) | |
|---|---|---|---|---|
| | HBond1 | HBond2 | HBond1 | HBond2 |
| NMR | 152.28 | 156.66 | 3.30 | 3.05 |
| Sim1 | $146.75 \pm 9.17$ | $119.81 \pm 20.06$ | $4.46 \pm 0.27$ | $3.43 \pm 0.63$ |
| Sim2 | $135.86 \pm 13.18$ | $122.68 \pm 15.48$ | $3.00 \pm 0.17$ | $3.46 \pm 0.25$ |

On the other hand,, Sim2 retains the experimentally observed bidentate fashion of Glu11 (see Fig. 4b and Tab. 3). From Tab. 4 one can see that both HBond1 and HBond2 are kept by Sim2. Thus the irregular $3_{10}$ helix can be observed in Sim2 (see Fig. 4b), which is consistent with the experimental observation. As a result, Sim2 can provide satisfied coordination bond lengths with NMR data.

### 3.4 Solvent exposure of residues

Here, the solvent exposure of a residue (SE-Res) is specified by the solvent accessible surface area of the residue. As illustrated in Fig. 5, expect for the residue Glu3, to which either Sim1 or Sim2 provides overestimated SE-Res, Sim2 in general shows a better tendency of SE-Res over the a. a. residue sequence, because the mean absolute/relative error of SE-Res over the 17 residues is 33.7/0.40 $Å^2$ for Sim1 and 22.8/0.29 $Å^2$ for Sim2. Notably, in order to capture the global error, the individual big relative errors for residue 14 (21.7 $Å^2$ for Sim1 and 7.7 $Å^2$ for Sim2) are not counted in.

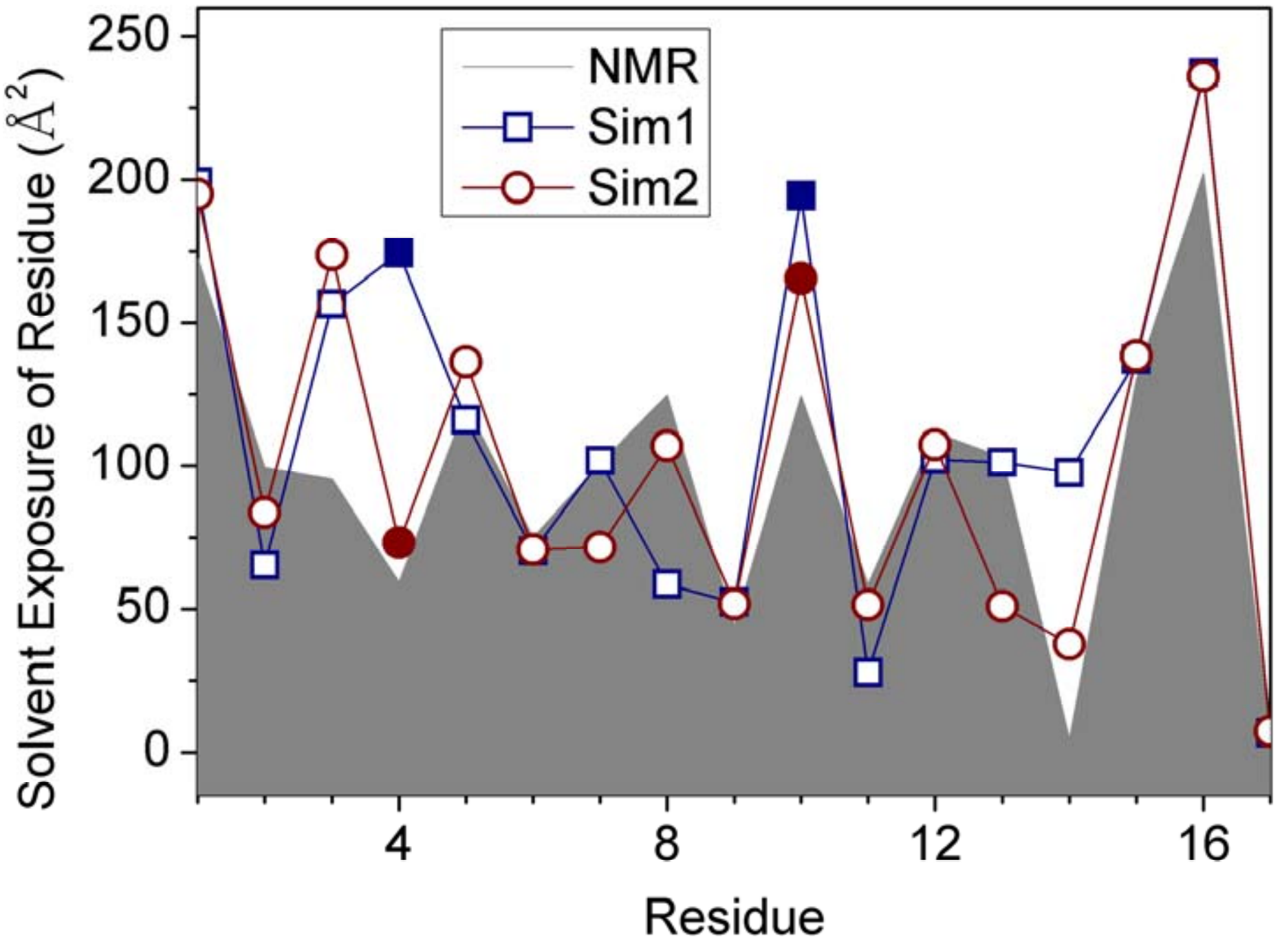

**Figure 5.** Solvent exposure of 17 residues in $A\beta_{1-16}$-Zn complex provided by Sim1 (open square) and Sim2 (open circle). Zinc is the 17$^{th}$ residue. Corner points of the shadow represent the solvent exposure of residues derived from the NMR structure. The results for residues Tyr10 and Phe4 are highlighted with solid square for Sim1 and solid circle for Sim2

Here, both Sim1 and Sim2 retain the four buried coordinating residues (His (6,13

and 14) and Glu11) due to the existing of zinc. However, comparing to the experimental values, the estimated values of SE-Res of Glu11 and His14 with Sim1 are worse than those with Sim2. It has been noted in [21] that, in contrast to the locations of residues Phe4 and Tyr10 in the apo $A\beta_{1-16}$ monomer (PDB ID: 1ZE7), residue Phe4 is located in the inner core of the structures and Tyr10 is excluded from the coordination sphere and systematically locates on the surface opposed to the zinc ion. As highlighted in Fig. 5, the results for Tyr10 and Phe4 are plotted with solid square for Sim1 and solid circle for Sim2. Both Sim1 and Sim2 are able to replicate the experimentally observed behavior of Tyr10. Nevertheless, Sim1 overestimates the SE-Res of Phe4 which leads to the Phe4, like in apo $A\beta_{1-16}$, staying on the surface of the peptide [21]. On the other hand, for Sim2, in which the polarization effect has been considered for the whole domain, the SE-Res of Phe4 is consistent with the experimental data.

## 3.5 Ligands exchange

There is not any explicit water molecule involved in the implicit solvent simulations, so it's difficult to judge whether our model (Sim2) is able to capture the experimentally observed tetrahedral coordination mode once explicit water molecules are presented. We found that at least it's negative for the explicit solvent simulation with the traditional force field using vdW2, because zinc is found to be hexocoordinated with the four residues mentioned above and another two water molecules.

Consequently, we design a simple hybrid solvent model that the bulk solvent is implicitly presented with the two coordinating water molecules explicitly described. Here, the explicit water molecules are modeled with TIP3P model and force field parameters come from CHARMM22 [46]. The starting structure is the equilibrant structure derived from an explicit solvent simulation with the traditional force field, in which vdW2 is used. Note that in the starting structure the carboxylate of Glu11 monodentately binds to zinc. Following the scheme suggested by Sakharov and Lim [29], the induced polarization effect and the charge transfer effect on water molecules are ignored in the current implement.

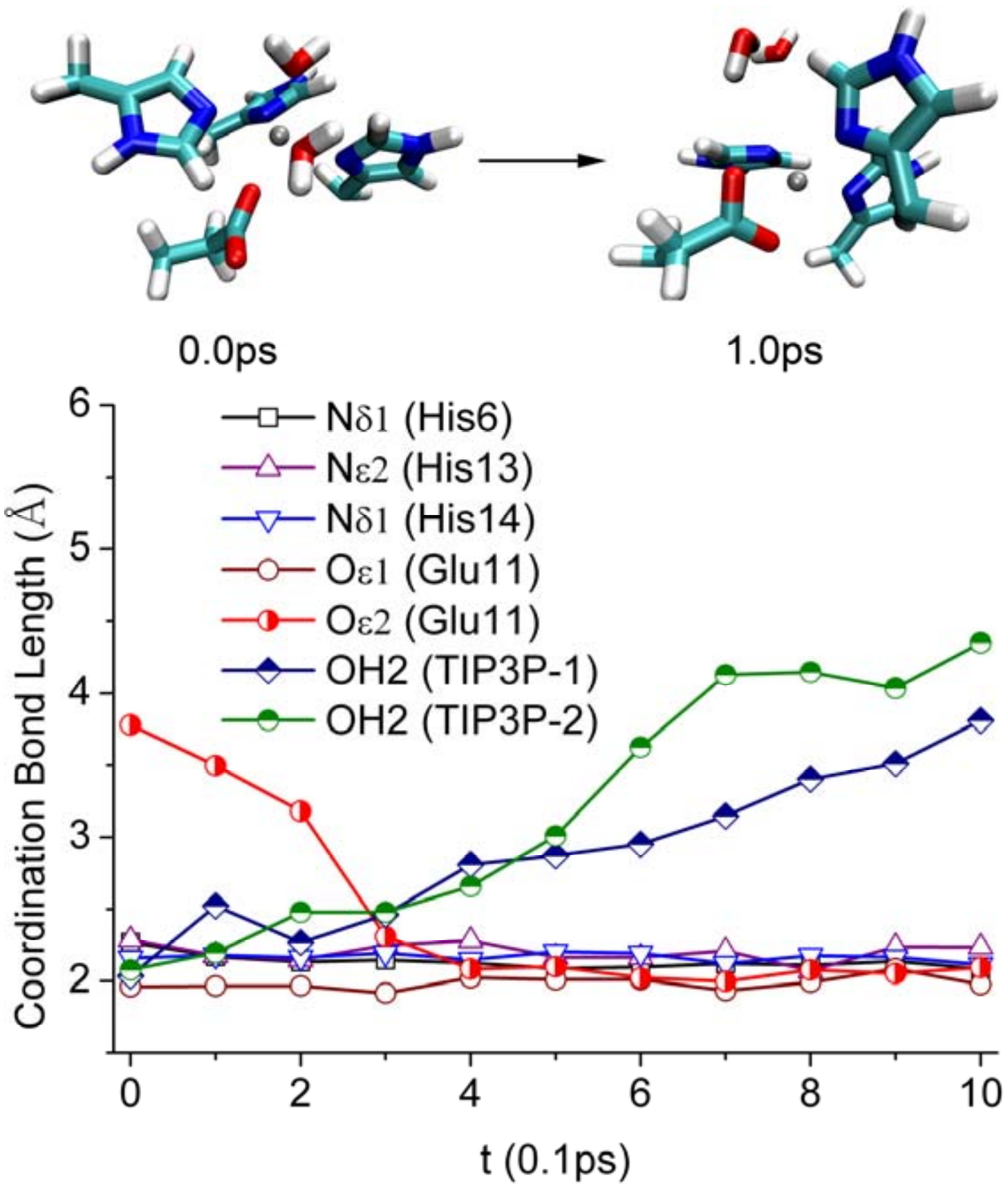

**Figure 6.** Coordination bond length varying as a function of simulation time. Seven atoms are involved, including $N\delta 1$ (His6) given by open square, $N\varepsilon 2$ (His13) by up triangle, $N\delta 1$ (His14) by down triangle, $O\varepsilon 1$ (Glu11) by open circle, $O\varepsilon 2$ (Glu11) by red half-open circle, $OH2$ (TIP3P-1) by half-open square and $OH2$ (TIP3P-2) by green half-open circle, respectively. Snapshots of $A\beta_{1-16}$-Zn at the time point 0.0 ps and 1.0 ps are extracted from the MD simulation in the hybrid solvent. For the $A\beta_{1-16}$, only the coordinating sidechains are displayed. The gray ball is zinc. C, N, O and H atoms are colored in green, blue, red and white, respectively.

The coordination bond lengths for seven atoms, including $N\delta 1$ (His6), $N\varepsilon 2$ (His13), $N\delta 1$ (His14), $O\varepsilon 1$ (Glu11), $O\varepsilon 2$ (Glu11), $OH2$ (TIP3P-1) and $OH2$ (TIP3P-2), are presented in Fig. 6 as a function of simulation time. The results show that at the beginning the bond lengths concerned with the oxygens on the two explicit water molecules increase and simultaneously the bond length of $O\varepsilon 2$ (Glu11)-Zn reduces. The exchanging process ends at 0.4ps and from then on the new $O\varepsilon 2$ (Glu11)-Zn bond maintains. As to another four atoms not involved in ligand exchanging, their bond lengths in average show nearly no change all through the simulation. We also find that if the traditional force field is used, there is not any exchange of ligands observed, which demonstrates that implicit solvent does not influence the coordinating configuration generated in explicit solvent.

## 3.6 MD simulation of rat $A\beta_{1-16}$ dimmer complexed with zinc

The NMR solution structure of rat $A\beta_{1-16}$ dimmer upon zinc binding (rat $2A\beta_{1-16}$-Zn) (PDB code: 2LI9) is determined recently to explore the mechanism of rats' resistance to pathogentic $A\beta$ aggregation in AD [65]. The NMR measurements of this complex were carried out also at 278K. In the rat $2A\beta_{1-16}$-Zn complex, zinc is tetrahedrally coordinated with four histidines. Both rat $A\beta_{1-16}$ monomers contribute two histidines (His6 and His14). Obviously, the zinc binding sites is catalytic. The charge transfer parameters listed in Tab. 1 are employed also for the system. According to the scheme of multiscale dipoles, the sidechains of the four histidines use atom-based dipoles, the corresponding four backbone segments adopt backbone-based dipoles and other 28 residues employ residue-based dipoles. Similarly, we carry out another two simulations for 4ns with respect to this system. One simulation (rSim1) follows the scheme by Sakharov and Lim and the other (rSim2) uses our model.

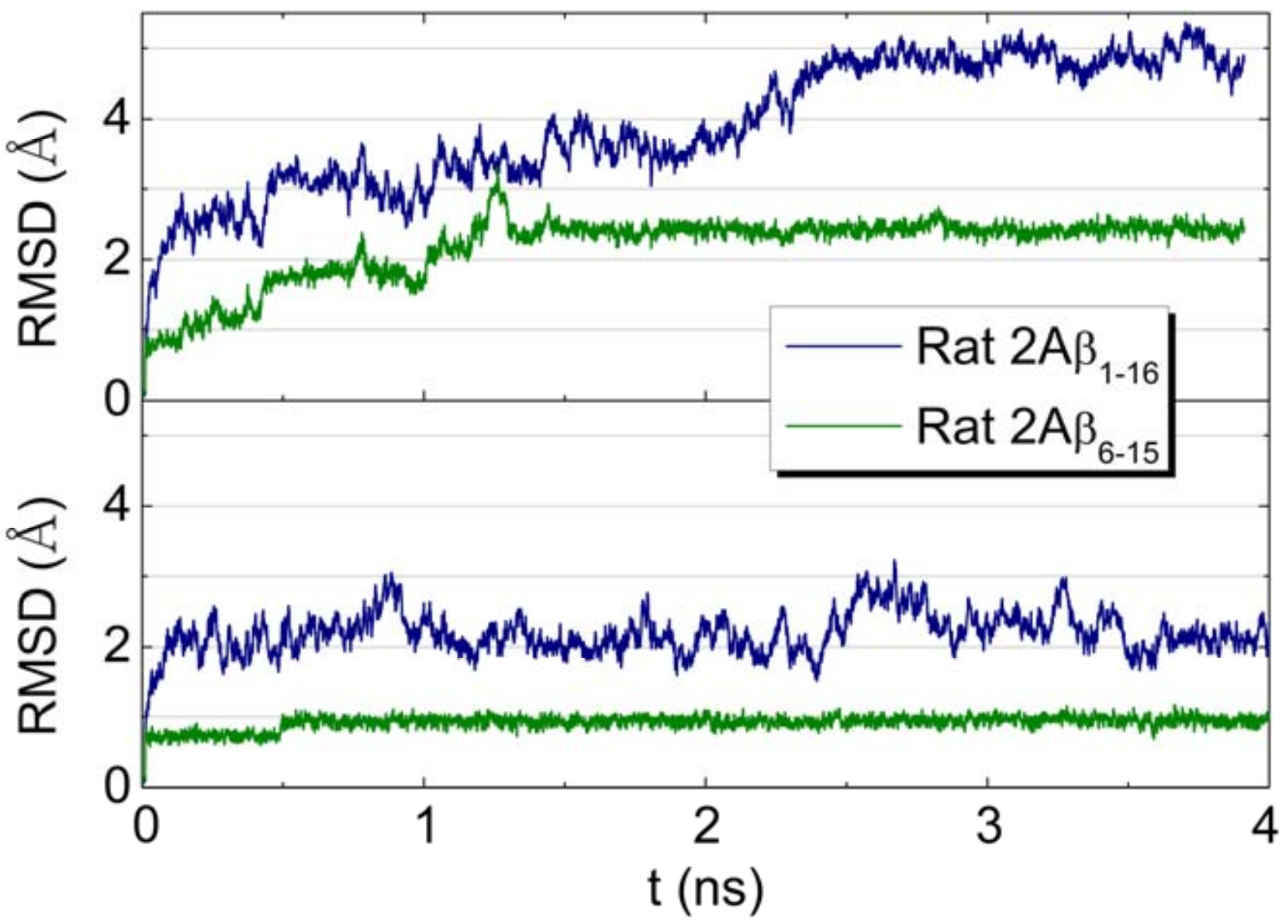

The RMSD of the whole dimmer (rat 2 $A\beta_{1-16}$) as well as the zinc binding domain (rat 2 $A\beta_{6-15}$) is calculated as a function of time. The RMSD of zinc binding center, including four ligating atoms and zinc, is also calculated. As is shown in Fig. 7, rSim1 is not able to reproduce the NMR structures of either rat 2 $A\beta_{1-16}$ or rat 2 $A\beta_{6-14}$. In contrast, the structures, especially for the 2 $A\beta_{6-14}$, provided by rSim2 are in agreement with NMR structures.

## 4. Discussion

In the paper, we propose an all-atom MD to discuss the complex $A\beta_{1-16}$-Zn with the induced polarization effect and charge transfer effect. As a comparison, another model which exactly follows Sakharov and Lim's method [29] is also discussed. From the results, we show that although the model Sim1 proposed by Sakharov and Lim is appropriate to the proteins with structural-type zinc binding sites [29], whereas it is subjected to three deficiencies in describing $A\beta_{1-16}$-Zn structure: (1) N terminus of $A\beta_{1-16}$ is not well stabilized and the irregular $3_{10}$ helix in the C terminus of $A\beta_{1-16}$ is not found; (2) zinc binding site is over compact and Glu11 prefers monodentate binding to zinc; and (3) Phe4 locates on the surface of $A\beta_{1-16}$. Nevertheless, all three deficiencies have been solved with our model Sim2.

From the RMSD of zinc binding center, we show that, our model can provide reasonable coordination structure. There are two factors responsible for this result. Firstly, Sim2 uses vdW2, in which the radius of zinc is bigger than the one in vdW1 adopted by Sim1. Such a bigger radius of zinc straightly looses the zinc binding site. Secondly, zinc-induced rearrangement of electrons on the coordinating sidechains is considered to be important. Taking His13 as an example, from the QM computation we find that even though zinc accepts charge from the coordinating atom $N\varepsilon 2$, the net charge on $N\varepsilon 2$ still increases, rather than reduces. The increasing part is contributed by the rest atoms of the imidzaole ring. In Ref. [30], as zinc get closing to the sidechain of His, the induced polarization of the whole imidzaole is presented by increasing the charge of the coordinating atom $N\varepsilon 2$ with the decreasing charge on the rest atoms in the imidazole ring. The model Sim1 modifies only the polarizability of the ligating atoms that directly binds to zinc. While in the current paper, we reset the atomic polarizabilities on the ligating sidechains to present the electron-delocalization effect, such as the increased polarizability of the coordinating atom $N\varepsilon 2$ and the reduced polarizability of the rest atoms in the imidazole ring.

Regarding the RMSD results of the whole $A\beta_{1-16}$ peptide, we suggest that polarization force induced by zinc in the domains outside the zinc binding site should be taken into account so as to stabilize the N-terminal region of $A\beta_{1-16}$ which is disordered in apo $A\beta_{1-16}$. A significant amount of charge ($\approx 0.7e$) [29] transferred from coordinating atoms to zinc in the structural-type zinc binding center leads to obvious reduce of charge on zinc and subsequently weakens the polarization effect of zinc accordingly. Therefore, it is unnecessary to globally consider the induction effects of zinc on zinc proteins with structural-type zinc binding centers. On the other hand, the charge transfer effect plays less important roles in the interaction between $A\beta_{1-16}$ and zinc, because much less charge ($\approx 0.4e$) is transferred from coordinating atoms to zinc. As a result, the induction effect of zinc out of the zinc binding site should not be ignored for $A\beta_{1-16}$-Zn.

So far as we know, we are the first to capture the bidentate mode of Glu11 bound to zinc in $A\beta_{1-16}$-Zn complex with molecular dynamic simulation based on the model without any bond or angle constraint. In fact, from the optimized structure of the complex $[Zn^{2+}(IMI)_3HCOO^-]^{1+}$, the carboxylate forms asymmetrical ZN-O bonds with zinc (1.98A and 2.48A), which is consistent with the results derived from the previous QM studies in Ref. [34,66]. It's been suggested that the bidentate binding fashion characterized by weaker Zn-O bond strength and similar Zn-O bond length observed in proteins depend not only on the Zn-O interactions but also on other interactions among ligands [67] under a certain protein environment. We found that, with the Sim2, $O\varepsilon 1$ (Glu11) involves in two hydrogen bonds with $H\beta 2$-$C\beta$ ( His14) and H-N (Glu11) respectively, while $O\varepsilon 2$ (Glu11) keeps strong interaction with $H\varepsilon 1$ (HIS13) (data not shown). These extra interactions are in agreement with the NMR structure. As a consequence, we are able to capture the bidentate binding mode. Differently, the use of smaller radius of zinc in Sim1 directly leads to the $O\varepsilon 1$ (Glu11) being excluded due to the limit in the space of the zinc binding center.

One obvious change in secondary structure between apo $A\beta_{1-16}$ and $A\beta_{1-16}$-Zn is the irregular $3_{10}$ helix formed in the C-terminal region of $A\beta_{1-16}$ as zinc binding to it. Our simulation successfully replicates this secondary structure because we are able to obtain the two critical hydrogen bonds in the helix. Apart from the secondary structure change, another important phenomenon is the reorientation of peptide residues due to the binding of zinc.

It has been suggested that one of the possible zinc-induced aggregating mechanisms is that zinc induces the conformational change of $A\beta$ to facilitate aggregation [8]. For example, zinc induces exposing of hydrophobic groups, which promotes $A\beta$ aggregation in aqueous solvent. It's easy to understand that positively charged zinc ion in the core of the peptide draws the hydrophilic or negatively charged groups into the core of the peptide due to the strong interaction between point charge of zinc and permanent dipoles or negative charge of a. a. residues, while hydrophobic or positively charged groups would be excluded and exposed to the solvent.

However, for $A\beta_{1-16}$ upon zinc binding, it's difficult to explain why the nonpolar and neutral Phe4 locates in the core of the peptide occupied primarily by zinc and those polar or negatively charged residues. One possible interpretation is that Phe4 involves in hydrophobic interactions with other nonpolar groups. However, there are only four nonpolar residues (Ala2, Phe4, Glu9 and Val12) in $A\beta_{1-16}$ peptide and no hydrophobic core composed by them is found in either $A\beta_{1-16}$ or $A\beta_{1-16}$-Zn. Therefore, we suggest that the location of Phe4 is modulated mainly by zinc through the induced polarization force. Meanwhile, the solvent exposures of another three nonpolar residues provided by Sim2 are in good agreement with the experimental observation. Here, we suggest that the zinc-induced dipole on Phe4 forms considerable attractive interaction with zinc, which makes the Phe4 to stay in the core of the peptide. On the other hand, like the situation of Phe4 in the apo $A\beta_{1-16}$ monomer, Phe4 locates on the surface of the $A\beta_{1-16}$-Zn complex in Sim1 that ignores the polarization force on this a. a. residue.

In order to demonstrate the ability of our model to reproduce the tetrahedral coordination mode in the presence of explicit water molecules, we have carried out an additional MD simulation with a hybrid solvent. Our simulation result echoes the conclusion in Ref. [29] that the induced polarization effect combined with charge transfer effect is critical in avoiding unwanted water bound to zinc in the $A\beta_{1-16}$-Zn complex. As observed in Fig. 6, either explicit or implicit solvent simulation provides similar coordinating bond lengths for $N\delta 1$ (His6), $N\varepsilon 2$ (His13), $N\delta 1$ (His14), $O\varepsilon 1$ (Glu11). This result implies that individual coordinating bond length is predominated by vdW radius of zinc. As a result, vdW2 is essential to reproduce satisfied coordinating bond length for $A\beta_{1-16}$-zinc peptide.

## 5. Conclusion

The objective of this work is to propose an efficient and reliable scheme compatible with CHARMM22 force field to describe the interaction between zinc and $A\beta_{1-16}$-like peptide, especially the metal-induced polarization force. Here, the point dipole model has been applied to present the polarization interaction and the induced dipoles have three distinct scales according to their distances from zinc ion. Regarding the charge transfer effect, the amount of charge transferred from a ligating atom to zinc is linearly related with zinc-ligand distance. MD simulations with implicit solvent have been implemented by our self-developed MD program. The non-polar contribution of solvation free energy has been parameterized in this work so as to be consistent with the CHARMM22 force field. With our model, we have successfully obtained the zinc binding site in good agreement with the crystal structures and are the first to capture the Glu11 of $A\beta_{1-16}$ bound to zinc in a bidentate fashion without any bond or angle constraint. Moreover, we have obtained the stabilized N-terminal and C-terminal regions of $A\beta_{1-16}$ and the secondary structure in $A\beta_{1-16}$-Zn as observed in experiment. Finally, reasonable solvent exposure of $A\beta_{1-16}$ a. a. residue, which plays a key role in metal-coupled folding or oligomerization of $A\beta_{1-16}$-like peptide, has been yielded based on our model. These results indicate that, to model $A\beta_{1-16}$-Zn complex involved in Alzheimer's disease, (1) $R_{Zn}$ =1.09 Å is more appropriate as the radius of zinc in the framework of CHARMM22 force field in studying catalytic-type zinc binding sites and (2) polarization effect should be extended to the domains beyond the zinc binding site for $A\beta_{1-16}$-like peptide. From the simulations of rat $A\beta_{1-16}$ dimmer upon zinc binding, we suggest that our model can be applied to other $A\beta_{1-16}$-Zn like complexes.

# Acknowledgment

We acknowledge support from the National Natural Science Foundation of China under grant 30970970, the China National Funds for Distinguished Young Scientists under grant 11125419 and the Funds for the Leading Talents of Fujian Province. Computational supports from the Key Laboratory for Chemical Biology of Fujian Province, Xiamen University and Xiamen Super Computing Center are gratefully acknowledged.

# Supplementary Material.

**Table 1.** LCPO and LCPO/NLR parameters. The four parameters (P1, P2, P3, P4) employed in the LCPO as well as LCPO/NLR method are shown. They are derived from a database of 61 Zn proteins containing by multiple linear regression fitting by using equation 16, with solvent surface accessible areas obtained with the analytical method available within the TINKER package (http://dasher.wustl.edu/tinker/) as independent variable. The number of atoms employed in the fitting for each atom type is shown. In our implementation, some atom types, such as hydrogen, were further

distinguished by their van der Waals radii derived from CHARMM22 force field in order to improve the fittings.

Table 1.

| Atom (Hybrid) | No. of Bonded Neighbors[a] | NO. of atoms | vdW radius[b](Å) | Methods | P1 | P2 | P3 | P4 |
|---|---|---|---|---|---|---|---|---|
| C(sp2) | 2 | 653 | 1.8 | LCPO<br>LCPO/NLR | 0.55465<br>0.35797 | -0.07591<br>-0.11224 | 4.78574e-4<br>0.00626 | 5.19636e-5<br>4.34434e-5 |
| C(sp2) | 2 | 3720 | 1.9924 | LCPO<br>LCPO/NLR | 0.14643<br>0.17375 | -0.02526<br>-0.05138 | 4.27957e-5<br>0.0024 | 1.13126e-5<br>2.43628e-5 |
| C(sp2) | 3 | 13505 | 1.8,1.9924, 2.000 | LCPO<br>LCPO/NLR | 0.09217<br>0.09856 | -0.01826<br>-0.03136 | 4.39035e-5<br>0.0024 | 8.74947e-6<br>4.01782e-6 |
| C(sp3) | 1 | 5228 | 2.06 | LCPO<br>LCPO/NLR | 0.23394<br>0.24985 | -0.04324<br>-0.07611 | 5.72052e-5<br>0.00313 | 2.09284e-5<br>4.22179e-5 |
| C(sp3) | 2 | 12803 | 2.175 | LCPO<br>LCPO/NLR | 0.19935<br>0.2015 | -0.04418<br>-0.06821 | 6.08465e-5<br>0.00296 | 2.36207e-5<br>3.84186e-5 |
| C(sp3) | 3 | 11565 | 2.275 | LCPO<br>LCPO/NLR | 0.0833<br>0.08026 | -0.01628<br>-0.02517 | 4.84336e-5<br>0.00102 | 7.37073e-6<br>1.32528e-5 |
| H(sp3) | 1 | 52592 | 1.32 | LCPO<br>LCPO/NLR | 0.35592<br>0.45365 | -0.06112<br>-0.12906 | 3.5963e-5<br>0.00499 | 3.03026e-5<br>8.40192e-5 |
| H(sp3) | 1 | 3719 | 1.3582 | LCPO<br>LCPO/NLR | 0.46783<br>0.52653 | -0.08454<br>-0.15076 | -9.26511e-5<br>0.00443 | 4.83759e-5<br>1.2189e-4 |
| H(sp3) | 1 | 48 | 1.468 | LCPO<br>LCPO/NLR | 0.64212<br>0.68591 | -0.13494<br>-0.2247 | 1.33799e-4<br>0.00954 | 7.72179e-5<br>1.64581e-5 |
| N(sp2) | 1 | 1702 | 1.85 | LCPO<br>LCPO/NLR | 0.70315<br>0.72133 | -0.15994<br>-0.25136 | 3.14806e-4<br>0.0082 | 8.95011e-5<br>2.28935e-4 |
| N(sp2) | 2 | 10293 | 1.85 | LCPO<br>LCPO/NLR | 0.29462<br>0.35526 | -0.05632<br>-0.11003 | 7.65035e-5<br>0.00596 | 2.76667e-5<br>4.90476e-5 |
| N(sp2) | 3 | 468 | 1.85 | LCPO | 0.01626 | -0.00254 | 3.90873e-5 | 1.95677e-7 |
| | | | | LCPO/NLR | 0.04892 | -0.01098 | 0.00261 | -2.4607e-5 |
| N(sp3) | 1 | 717 | 1.85 | LCPO | 0.77392 | -0.17982 | 1.32462e-4 | 1.08084e-4 |
| | | | | LCPO/NLR | 0.77641 | -0.25459 | 0.0082 | 2.03257e-4 |
| O(sp2) | 1 | 12886 | 1.7 | LCPO | 0.59681 | -0.11112 | -3.35408e-4 | 6.72611e-5 |
| | | | | LCPO/NLR | 0.6376 | -0.18472 | 0.0057 | 1.38405e-4 |
| O(sp3) | 1 | 1428 | 1.77 | LCPO | 0.63278 | -0.12674 | 2.23384e-4 | 6.49087e-5 |
| | | | | LCPO/NLR | 0.66214 | -0.19791 | 0.00754 | 1.30401e-4 |
| S(sp3) | 1 | 221 | 2.2 | LCPO | 0.59019 | -0.11898 | 2.69179e-5 | 6.70281e-5 |
| | | | | LCPO/NLR | 0.57959 | -0.17974 | 0.0057 | 1.52524e-4 |
| S(sp3) | 2 | 18 | 1.975 | LCPO | 0.51241 | -0.09622 | -2.70732e-4 | 5.80067e-5 |
| | | | | LCPO/NLR | 0.48538 | -0.14182 | 0.00358 | 1.20227e-4 |
| S(sp3) | 2 | 170 | 2.0 | LCPO | 0.53736 | -0.11554 | 3.65727e-4 | 6.00726e-5 |
| | | | | LCPO/NLR | 0.53051 | -0.16056 | 0.00315 | 1.59467e-4 |
| Zn(sp3) | 4 | 113 | 0.88[c] | LCPO | 0.66025 | -0.15985 | -8.06554e-4 | 1.44219e-4 |
| | | | | LCPO/NLR | 0.65348 | -0.24569 | 0.0046 | 3.72103e-4 |

a. The number of bonded neighbors that are not hydrogen atoms.
b. The van der Waals radii are derived from parameter document of CHARMM22 force field.
c. The radius of zinc is small that it contributes little to the total screening areas of its neighbors. Therefore the parameter set is also suit to the other zinc radius (1.09 Å),